\documentclass[conference]{IEEEtran}
\usepackage{algorithm}
\usepackage{algorithmic}
\usepackage{amsthm}
\usepackage[pdftex]{graphicx}
\usepackage[top=20mm, left=19mm,right=19mm,bottom=20mm, papersize={210mm, 297mm}]{geometry}
  
\ifCLASSINFOpdf
\else
\fi
\begin{document}
\title{Search Result Clustering via Randomized Partitioning of Query-Induced Subgraphs}
\author{\IEEEauthorblockN{Aleksandar Bradic}
\IEEEauthorblockA{Faculty of Electrical Engineering, Belgrade\\
Email: abradic@acm.org}
}
\maketitle

\begin{abstract}
In this paper, we present an approach to search result clustering, using partitioning of underlying link graph. We define the notion of "query-induced subgraph" and formulate the problem of search result clustering as a problem of efficient partitioning of given subgraph into topic-related clusters. Also, we propose a novel algorithm for approximative partitioning of such graph, which results in cluster quality comparable to the one obtained by deterministic algorithms, while operating in more efficient computation time, suitable for practical implementations. Finally, we present a practical clustering search engine developed as a part of this research and use it to get results about real-world performance of proposed concepts.
\end{abstract}

\IEEEpeerreviewmaketitle

\section{Introduction}

Efficient representation of search results poses a significant challenge for modern search engines. The widelly-accepted score-based 
model [1], although quite effective in the general case of search for the \textit{best} document matching the given query,  is usually insufficient in situations which require representation of larger set of relevant results. This is especially true in the case of clustering and exploratory search engines, which focus not only on representation of the relevance, but also the way the results are related and their organization into \textit{clusters} of related documents. 

Clustering based on document information content has been a well studied topic in Information Retrieval (\textit{IR}). Standard \textit{IR} clustering methods, based on the \textit{cluster hypothesis} [2], usually operate by calculating appropriate content-based relevance values and imposing certain \textit{similarity} metric, have been accepted by Search Engine community and implemented in a number of real-world clustering engines (\textit{Vivisimo}, \textit{Carrot Clustering Engine}, \textit{Mooter}, \textit{Clusty}). 
Still, we can observe that clustering Web data in this manner fails to capture the essential component of Web documents, which is the \textit{hyperlink} information, reflected in \textit{link graph}, which describes the explicit way in which the documents are connected. A lot of algorithms utilize this structure to extract information about document relevance (\textit{PageRank} [1]), and community structure (\textit{HITS} [3]). Great success of these algorithms, indicated the significance of link structure in Web data analysis, and suggested extension of such concept to other related problems, like the problem of \textit{community detection} [3], and \textit{Web data clustering} [4]. 
However, although there are significant results in the area of link-based Web data clusterings, implementing such algorithms in practical search engine still poses a significant challenge, primarly due to the fact that, unlike \textit{IR}-based methods, which operate on set of values \textit{precomputed} for each document, graph-based algorithms operate on dynamical query-dependent representation of entire link graph, which makes precomputation impossible and problem both computationally and space-intensive. As a result of this, currently there are no real-world clustering engines that implement search result clustering using the link-graph approach.

In this paper, we propose a relaxation of the problem of search result clustering from the problem of clustering the entire graph to the domain of \textit{query-induced sugraph}, representing a subgraph generated by given search query and show the validity of such proposal by determining that the essential structural properties of the entire graph are still preserved in given subgraph. Further, we propose a novel algorithm for approximative clustering of such subgraphs, which enables us more space and computationally efficient clustering, with variable margin of error, suitable for implementation in real-world search engines. Finally, we present a search engine called \textit{randomNode}, implemented as a part of this research, which demonstrates usability of proposed concepts in real-world application.
 
\section{Related Work}

In [5], authors perform the first analysis of the general structure of the Web, and determine that node degree distribution follows a simple power-law of the form $k^{-\theta}$, with $\theta = 2.1$ for in-degree and $\theta = 2.7$, for out-degree. In [6], a single subset of Web Graph is analyzed - the Web of a single country (\textit{Web of Spain}) and similar distribution is observed, with  $\theta = 2.11$ for in-degree and $\theta = 2.84$ for out-degree, validating the scale-free structure of the Web Graph and indicating that the link distribution is invariant to the change of scale (we use this idea in proposing the concept of \textit{query-induced subgraphs}). Complete statistical analysis of topic-related link graphs, generated in social networks, is given in [7]. Authors observe the power law distribution of node degrees and propose power law based on truncated-log-normal hypothesis. Finally, paper [8] gives a complete description of methods for estimating power-law distribution parameters from empirical data.

General overview of graph clustering algorithms and appropriate metrics for determining cluster quality is given in [9].  Efficient graph clustering algorithms vary in computational complexity from $O(n^3)$, in the case of \textit{recursive partitioning}, to  $O(nlogn)$ in the case of \textit{multilevel clustering} algorithm described in [10]. However, all of given algorithms operate in $O(n^2)$ space complexity, as they require availability of entire graph representation, making it hard for implementation on the scale of $n$ found in practical problems.

\section{Query-Induced Subgraphs}

\subsection{Definition}
Let the \textit{hyperlink graph} be a graph $G=(V,E)$, where $V$ is a set of vertices representing all the documents in the search engine index, and $E$ is a set of edges representing \textit{hyperlinks} between all the documents. 

We define \textbf{Query-Induced Subgraph} as a graph $G_q$ = $(V_q,E_q)$, where $V_q \subset V$ is a set of all results matching the given query $q$ and $E_q \subset E$ set of all edges between vertices from the set $V_q$. In practice, given subgraph ($G_q \subset G$) represents the hyperlink graph created from $G$ by keeping only the documents matching given query and hyperlinks between the documents in resulting set. We define \textit{node degree} as number of links (both inlinks and outlinks) for each node, and treat it as a measure of information content contained in link data. Our goal is to show that the \textit{node degree} in the given \textit{query-induced subgraph}, preserves the same distribution as in the entire graph (anticipated by the general assumption about \textit{scale-free} structure of Web and social networks [5]).

\subsection{Properties}

In order to validate the given assumption about degree distributions, we analyze the dataset obtained as a part of \textit{randomNode} clustering engine. Given dataset consists of data about 1.1 million nodes (representing the subset of \textit{.yu} Web), generated by calculating inlink degrees for resulting sets of 1000 top-frequency queries in \textit{randomNode} clustering engine. We analyze the distribution of inlink degrees for both full graph and induced subgraphs obtained for each of given queries and test the hypothesis that both graphs have distribution, commonly found in Internet and social networks [7] - a power law distribution with $\beta$ and $x_{min}$ parameters, and probability density function of the form : 
\begin{equation}
p(x;\beta,x_{min})=\frac{x^{-\beta}}{\zeta(\beta,x_{min})}
\end{equation}
where $\zeta(\beta,x_{min})$, represents the generalized zeta function
$\zeta(\beta,x_{min})=\sum_{n=0}^{\infty} (n+x_{min})^{- \beta}$.

We use the method of Maximum Likelihood (ML) for estimation of distribution parameters, as described in [8]. The approximate expression for MLE estimator of $\beta$ parameter is given by :
\begin{equation}
\hat{\beta} \equiv 1+ n \biggl{[} \sum_{i=1}^{\infty}
\ln{\frac{x_i}{x_{min} - \frac{1}{2} }} \biggl{]}^{-1}
\end{equation}

where $x_{min}$, represents the lower bound on the power law behavior. 

Figure 1 shows both the cummulative distribution function (\textit{cdf}) of node degrees in entire graph (full line) and in query-induced subgraphs (dotted line), obtained from the given dataset, as well as the \textit{cdf} of fitted power law distribution.
Estimated values for the $\beta$  using given procedures are shown in \textit{Table I}, with goodness of estimation given in terms of standard error. Given error values are in acceptable regions, confirming the hypothesis that the inlink distribution observed in given dataset can indeed be characterized by power-law distribution of the form given in formula \textit{(1)}.  

\begin{figure}[h]
 \scalebox{0.26}            
{ \includegraphics{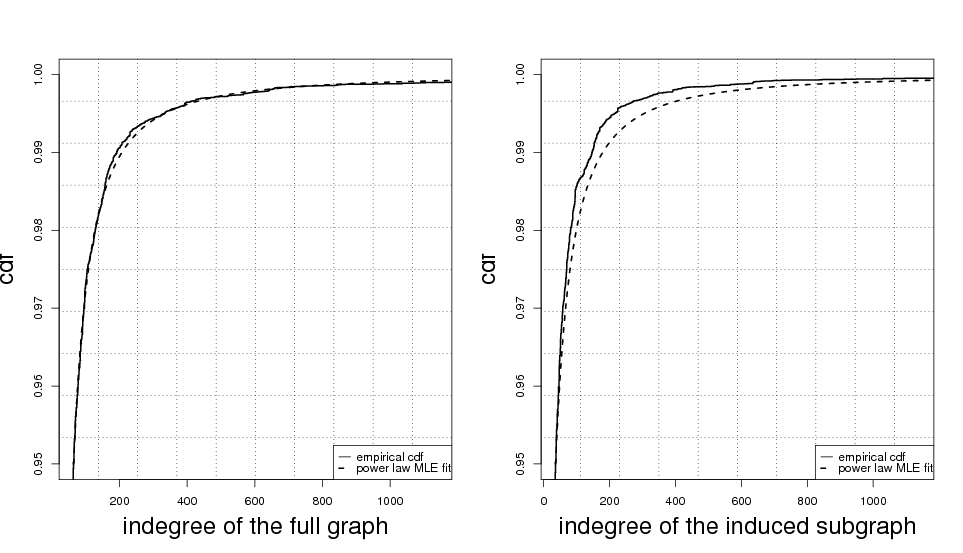}}
\end{figure}

\textit{Figure I : full graph and query-induced subgraph link degree distribution}\\

\begin{table}[ht] 
\caption{Link Distribution Power Law Fit}  
\centering       
\begin{tabular}{c c c c c}   
\hline\hline                         
& median & mean & $\hat{\beta}$ & std.error\\ [0.5ex]  
 
\hline                     
full graph & 3.00 & 17.96 & 2.500576 & 0.001184400\\   
induced subgraph & 1.00 & 9.98  & 2.533536 & 0.001531097 \\ [1ex]       
\hline      
\end{tabular} 
\label{table:fitstats}   
\end{table}

Finally, from Table I we observe estimated values of  $\beta =  2.500576$ for full graph and $\beta = 2.533536$ for induced subgraph, which validates the proposed concept of scale-invariance of graph structure. This further indicates that the essential graph properties (high-degree "authoritative" nodes [3] and random walk convergence properties [4]), existing in the entire graph, are still preserved in the query-induced subgraph.  Hence, we can reduce the dimension of \textit{search result clustering} problem, by restating it as a problem of clustering the \textit{query-induced subgraph} $G_q$, corresponding to the given query $q$. Such problem relaxation enables us to perform computation in much efficient manner, while still preserving essential information contained in the link structure.

\section{Algorithm for fast clustering using random walks on power-law graphs}

\subsection{Description}
We propose an algorithm for graph clustering using random walks on directed power-law graphs. The algorithm operates by performing a number of independent random walks on the link graph and attempts to exploit the specific structure of common power-law graphs in order to bound the average walk length. For each walk, we record a number of times each node was visited, and obtain partial sets, each containing the nodes visited during the walk and appropriate visit counts. Finally, we use that info in order to perform the \textit{merge} stage of the algorithm, in which we use \textit{pivot} nodes (nodes with maximum visit counts), in order to merge the given partial sets into a number of final sets, representing the \textit{cluster set} for a given graph.

\subsection{Algorithm}

Let the $G(V,E)$ be the connected, directed graph with $|V|=N$ and $|E|=m$. By \textit{random walk on graph}, we assume Markov chain $M_g$, where $V$ represents the set of \textit{states} of the chain and $P=[p_{ij}]$ is a stochastic matrix, with $p_{ij}$ representing transitional probability for any two states ${i,j \in V}$, given by :

\begin{equation}
P_{i j}  = \left\{\begin{array}{cl}
     \frac{1}{d(i)}, & \ \ {\rm if }  \ \   \exists (i,j) | (i
\rightarrow j) \in E \\
     0, & \ \ {\rm if } \ \ {\rm otherwise}
      \end{array}\right.
\end{equation}

and $d(i)$ represents the outdegree of a vertex $i$.

We define \textit{stationary distribution} of a Markov chain $M_g$ corresponding to a given walk on graph $G$, as a probability distribution $\bar{\pi}$, such that $\bar{\pi} = \bar{\pi} * P$, were each entry $\bar{\pi_i}$ is proportional to the amount of time walk will spend in a given node. Such distribution is often used as a measure of \textit{importance} of given node $i$. In the undirected case, the random walk on the graph converges to the stationary distribution [1], as well as in the case of directed strongly connected graph [12]. Allthough this does not hold for the general case of arbitrary walks on power law graphs, it does hold for the case of \textit{strongly connected components} of such graph, which are shown to exist in the general case of power law graphs [11]. Additionally, we define the \textit{stopping state} of random walk on directed graph as a state corresponding to the \textit{terminating} node, that is node $u$ such that $\not \exists v \in V | P_{uv} > 0$. We define the \textit{stopping time} of the walk as a number of steps of $M_g$ it takes for a chain to reach the \textit{stopping state}.

\begin{algorithm}[H]
  \caption{Random Walk Clustering}
  \label{alg1}
  \begin{algorithmic}
    \REQUIRE Graph $G(V,E)$ and approximation factor $K$ , where $|V| = N$, $k \in [0,1]$ and $K=k*N$

   \STATE \textit{WALK phase}:
    \STATE $i \leftarrow 0$
    \WHILE{$i \leq K$}
    \STATE $s \leftarrow rand(1,N)$
    \WHILE{$s \neq 0$}
    \IF{$\not\exists s | s \in w_i $}
    \STATE $w_i \leftarrow (s,1)$
    \ENDIF
    \STATE $s(w_i) \leftarrow s(w_i) + 1$
    \STATE $s \leftarrow rand(adj) ; v \in adj | \exists(s \rightarrow v) \in E$
    \IF{$adj = \{\}$}
    \STATE $s \leftarrow 0$
    \ENDIF
    \ENDWHILE
       \ENDWHILE
    \STATE \textit{we get the walk set} $W = (w_1....w_k)$

    \STATE    
    \STATE \textit{MERGE phase}:
    \STATE \textit{for each} $w_i \in W, i \in (1,K)$:
    \STATE \textit{for each node} $n \in w_i$:
    \IF{$\exists s \in w_k | |deg(s)-deg(n)| > T_{cm}$}
    \STATE \textit {we remove} \textbf{(cut)} \textit{node n from} $w_i$
    \ENDIF
    \IF{$\exists s \in w_k | |deg(s) - deg(n)| < T_{cm}$}
    \STATE \textit{we perform} \textbf{merge} of $w_i$ and $w_m$
    \ENDIF
    
    \RETURN $C = (w_1...w_m)$, $m \leq K$ - \textit{the final set of clusters in given graph} 

  \end{algorithmic}
\end{algorithm}

For the purpose of a given algorithm, we define \textit{stopping condition} for given walk either as a condition of process entering the \textit{stopping state}, or as a threshold value for the length of the walk. Due to the nature of the underlying graph, not every walk will enter the \textit{stopping state}, since the loops might occur, therefore we must define additional \textit{maximum walk length} $L$ (usualy of $O(N)$ order), which should prevent infinite loops, yet be large enough for the walk to capture the sufficient approximation of a distribution of node visit counts for given walk.

We perform the \textit{WALK} phase of the algorithm by selecting $K = k*N$ random nodes, where $k \in (0,1)$, represents the \textit{approximation constant} of the algorithm, and performing $K$ walks on graph $G$. Walks are performed untill they reach the stopping condition, either by entering the \textit{stopping state} or by hitting the \textit{maximum walk length}. 

Finally, in the \textit{MERGE} size, we sort walks by lenght, and internally by visit count, and iterate the result set by performing \textit{CUT} and \textit{MERGE} operations, interchangeably. If, for a given node, there is a walk having visit count \textit{significantly} greater than in the current walk, we remove it (\textit{CUT}) from given walk, whereas, if there is a walk having \textit{similar} visit count for a given walk, we perform \textit{MERGE} of two walks based on given (\textit{pivot}) node. In such manner, we hope to identify the \textit{key} (\textit{pivot}) nodes for every walk, and perform a join of two walks in case they share the key nodes. Additionally, by manipulating the threshold value for cut/merge ($T_{cm}$), we can efficiently manipulate the dimension of the clustering, balancing between cluster number and cluster size.

\subsection{Analysis}

In order to analyze given algorithm, we use results proved in [11], stating that for a class of power law graphs with $N$ nodes and exponents in range $\beta \in (2,3)$ (which correspond to the general case of Internet, social and citation networks, such as the dataset analyzed in this paper), average distance between any two is almost surely of order $O(loglog(N))$. In such a graph, it is guaranteed that there are more than zero terminating nodes, and the expected average distance between arbitrary node and given terminating node is of order $O(loglog(N))$. Therefore, we can determine that the expected average run length of the \textit{WALK} phase is of the $O(Nloglog(n))$ order. Additionally, such graphs contain the \textit{strongly connected component} of the size $n^{c/{loglog(n)}}$ [11], therefore, we define the $O(N)$ \textit{maximum walk length} in order to cover walks not hitting the terminating node. This finally results in $O(N^2)$ worst case time for a given algorithm and $O(Nloglog(N))$ expected average case time for the WALK phase and for a complete algorithm (the \textit{merge} phase can be implemented efficiently in $O(Nloglog(N))$ time).

However, although the worst case time of given algorithm is $O(N^2)$, both his average running time, and the fact that by reducing the problem to the induced subgraph, we operate on $N$ which represents the number of nodes matching the given query and is significantly smaller than the total number of nodes in search engine index. Additionally, given random walk implementation is much more space efficient, as it only requires storage of adjacency list for every node (of $O(Nlog(N))$ order) to perform random walks and get partial sets, as opposed to the matrix-based eigenvalue methods, which require $O(N^2)$ space for storage of the entire adjacency matrix.

\section{Results}

As a part of the research, and as a base for obtaining practical results, we have created a clustering search engine called $RandomNode$, accessible at \textit{http://www.randomnode.com}, which performs query-time clustering of search results by implementing the \textit{Random Walk Clustering} algorithm, proposed in section IV, implemented on top of the $Lucene$ search library. It operates on 1.1-million node dataset, represents a significant portion of $.yu$ web, generated by performing a crawl starting at the homepage of the Belgrade University (\textit{http://www.bg.ac.yu}). 

\begin{figure}[h]
\scalebox{0.22}            
{ \includegraphics{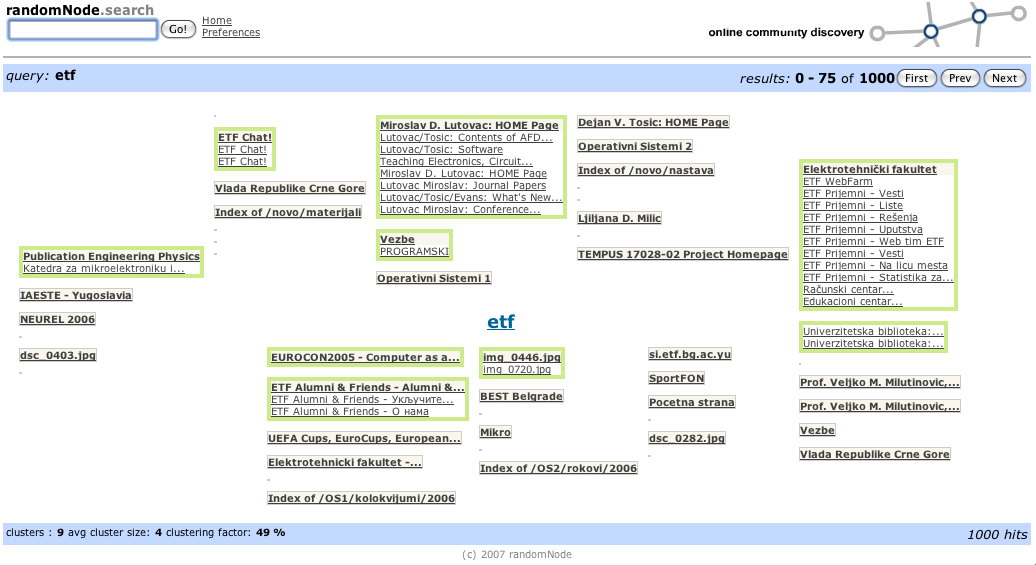}}
\end{figure}
\textit{Figure II : randomNode clustering engine}\\

We use the \textit{randomNode} clustering engine in order to analyze the impact of approximation factor $K$ on the performace of the proposed algorithm. We use the $coverage(C)$, of a graph clustering $C = (C_1 ... C_k)$, as as a measure of clustering quality, defined as: 
\begin{equation}
{coverage(C)} = {{m(c)}\over{m}} = { {m(C)}\over{m(C) + \bar{m}(C)} }
\end{equation}
where $m(C)$ represents the number of \textit{inter-cluster edges}, while $\bar{m}(C)$ represents a number of \textit{intra-cluster edges}. Optimal clustering should minimize the  $\bar{m}(C)$, as it represents the size of the \textit{cut} in the graph performed by given clustering.

\begin{figure}[h]
 \scalebox{0.40}            
{ \includegraphics{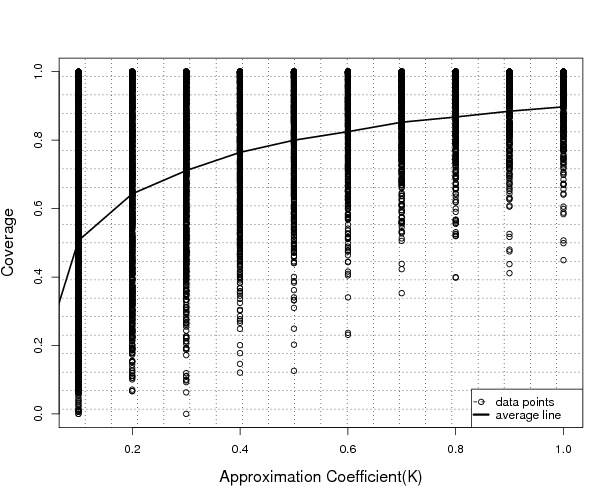}}
\end{figure}
\textit{Figure III : algorithm performance as function of approximation coefficient}\\

We perform analysis using \textit{randomNode} engine, by performing clustering on 1000 top-scoring keywords in given dataset, varying the approximation coefficient in the $(0.1,1.0)$ range with $0.1$ step and calculating the \textit{coverage} metric. The results are shown in Figure III, with scatterplot showing exact coverage values for each of each sample instance and the average coverage, given by the line segment. We observe that the coverage increases logarithmically with the approximation coefficient, which indicates that the algorithm can provide acceptable approximations, even for the small values of $K$.  Finally, we use the randomNode engine to extract a set of queries, shown in Table II, representing top-scoring clusters, both in terms of results and a cluster coverage, for a given subset of \textit{.yu} Web.
\begin{table}[ht] 
\caption{Top clusters in randomNode dataset}  
\centering       
\begin{tabular}{c c c c c c}  
\hline\hline                         
query & coverage & n.links & incluster & n.clusters & max size\\ [0.5ex]  
 
\hline                     
politika & 0.999 & 37473 & 37417  & 29 & 820\\    
pravda & 0.967 & 34688 & 33556 & 43 & 682\\        
rubrike & 0.995 & 33200 & 33053 & 13 & 817 \\        
shop & 0.967 & 29440 & 28482 & 88 & 549\\         
nekretnine & 0.989 & 28451 & 28157 & 30 & 535\\         
leasing & 0.988 & 28185 & 27847 & 35 & 272\\        
dekanat & 0.947 & 28783 & 27264 & 63 & 326\\        
banking & 0.965 & 26840 & 25916 & 120 & 211\\        
expo & 0.963 & 26456 & 24629 & 69 & 273 \\        
filologija & 0.976 & 23160 & 22609 & 39 & 625\\        
\hline     
\end{tabular} 
\label{table:topclusters}  
\end{table} 

\section*{Acknowledgment}

Thanks to prof. Veljko Milutinovic, who mentored this research as a part of my diploma thesis at the faculty of Electrical Engineering, Belgrade.

\end{document}